\documentclass[conference]{IEEEtran}
\IEEEoverridecommandlockouts
\usepackage{cite}
\usepackage{amsmath,amssymb,amsfonts}
\usepackage{algorithm, algorithmic}
\usepackage{graphicx}
\usepackage{textcomp}
\usepackage{xcolor}
\usepackage{subfigure}
\usepackage[hang,flushmargin]{footmisc}
\usepackage{lipsum}

\def\BibTeX{{\rm B\kern-.05em{\sc i\kern-.025em b}\kern-.08em
    T\kern-.1667em\lower.7ex\hbox{E}\kern-.125emX}}
\begin{document}

\title{Handling Distance Constraint in Movable Antenna Aided Systems: A General Optimization Framework\\
}

\author{\IEEEauthorblockN{Yichen Jin$^{1}$, Qingfeng Lin$^{1,2}$, Yang Li$^{2}$, and Yik-Chung Wu$^{1}$}
\IEEEauthorblockA{$^1$Department of Electrical and Electronic Engineering, The University of Hong Kong, Hong Kong}

\IEEEauthorblockA{$^2$Shenzhen Research Institute of Big Data, Shenzhen, China}

\IEEEauthorblockA{E-mail: \{\textcolor{black}{u3589542}, qflin, ycwu\}@eee.hku.hk, liyang@sribd.cn}

\thanks{The work of Yang Li was supported in part by the National Natural Science
Foundation of China (NSFC) under Grant 62101349 and Grant U23B2005, and also Guangdong Major Project of  Basic and Applied Basic Research (No.  2023B0303000001).}
}

\maketitle
%在maketile下加入这一段
\newcommand\blfootnote[1]{%
\begingroup
\renewcommand\thefootnote{}\footnote{#1}%
\addtocounter{footnote}{-1}%
\endgroup
}

% \blfootnote{The work of Yang Li was supported in part by the National Natural Science
% Foundation of China (NSFC) under Grant 62101349 and Grant U23B2005, and also Guangdong Major Project of  Basic and Applied Basic Research (No.  2023B0303000001).}

\begin{abstract} 

The movable antenna (MA) is a promising technology to exploit more spatial degrees of freedom for enhancing wireless system performance. However, the MA-aided system introduces the non-convex antenna distance constraints, which poses challenges in  the underlying optimization problems. To fill this gap, this paper proposes a general framework for optimizing the MA-aided system under the antenna distance constraints. Specifically, we separate the non-convex antenna distance constraints from the objective function by introducing auxiliary variables. Then, the resulting problem can be efficiently solved under the alternating optimization framework. For the subproblems with respect to the antenna position variables and auxiliary variables, the proposed algorithms are able to obtain at least stationary points without any approximations. To verify the effectiveness of the
proposed optimization framework, we present two case studies: capacity maximization and regularized zero-forcing precoding. Simulation results demonstrate the proposed optimization framework outperforms the existing baseline schemes under both cases.

\end{abstract}


\section{Introduction} \label{intro}

Driven by the explosive growth of wireless applications, including high-quality video streaming and virtual/augmented reality (VR/AR), the demand for high capacity in future sixth-generation (6G) communication networks has become increasingly apparent. To meet this requirement, the novel concept of movable antenna (MA)~\cite{zlp_magzine_ma, ma_icc}, also known as fluid antennas~\cite{kk_ma, kk_fl}, has been recently proposed. In contrast to the conventional multiple-input multiple-output (MIMO) systems with the fixed-position antenna (FPA), the MA can flexibly change its position, thus enabling more spatial degrees of freedom for enhancing the system performance.

Due to this unique benefit, the MA-aided systems have attracted significant attention. For example,~\cite{zlp_multiuser} leveraged the MAs to improve the channel capacity in multi-user communication systems. Moreover,~\cite{ma_channel} explored the channel estimation for the MA-aided system by utilizing the compressed sensing (CS) based approach, which offers a practical method for implementing MA technology. Furthermore, the MA technology has also been embedded in other communication scenarios, such as physical layer security~\cite{wuqingqing_secure} and integrated sensing and communication~\cite{isac_fl}. 

However, from the perspective of optimization, the MA-aided system introduces the non-convex antenna distance constraints, along with the possible non-convex objective function. These pose challenges in solving the underlying optimization problems.  Existing studies attempt different approaches to their specific optimization problems, e.g., successive convex approximation (SCA)~\cite{ma_icc}, and CS based approach~\cite{yang2024flexible}. However, they either require a convex approximation of the objective function and distance constraints, or introduce additional non-convex constraints (e.g., zero-norm constraint~\cite{yang2024flexible}), which leads to the performance loss.

To fill this gap, we aim to propose a general optimization framework for the MA-aided system under the non-convex antenna distance constraints. Specifically, to separate the non-convex antenna distance constraints from the objective function, we employ variable splitting by introducing the auxiliary variables. Then, we tackle the resulting problem via the alternating optimization. For the subproblems with respect to the antenna position variables and auxiliary variables, the proposed algorithms are able to obtain at least stationary points without employing any approximations or introducing additional constraints. We further demonstrate the proposed optimization framework using two typical examples: capacity maximization and regularized zero-forcing precoding, respectively. Numerical results show that the proposed framework outperforms state-of-the-art approaches under both studied cases. By providing a unified solution, this framework offers researchers a more streamlined and standardized approach to investigate MA based optimization problems.

\section{Distance Constraint in MA-Aided System}   \label{distance}

Consider such an MA-aided system, which
consists of one base station (BS) equipped with $M$ MAs, and $K$ $N$-antenna devices. Let $\mathbf{r}_m$ denote the position of the $m$-th ($m = 1, 2, \dots, M$) MA at the BS. Based on the analysis in~\cite{ma_icc}, the MA-aided systems generally introduce the following two additional constraints:
\begin{align} \label{1}
\mathbf{r}_m \in \mathcal{C}, ~~\forall m =1,2, \ldots, M, 
\end{align}
\begin{align} \label{2}
 \left\|\mathbf{r}_m-\mathbf{r}_l\right\|_2 \geq D, \quad  \forall m,l=1,2, \ldots, M, \quad m \neq l,
\end{align}
where $\mathcal{C}$ represents the given region for MAs to move and $D$ denotes the minimum distance between
each pair of antennas. Specifically, the first constraint restricts the MA's movement area, and the second constraint is referred to as
\textit{the antenna distance constraint} in order to avoid the coupling between the antennas in the given region $\mathcal{C}$~\cite{MIMO_cap_cha_for_MA}.

%In the next section, we will explain the challenges brought by the constraints~\eqref{1} and~\eqref{2} and then propose a general optimization framework for the MA-aided system.

\section{A General Optimization Framework for MA-Aided System}   \label{framework}
Consider the following optimization problem:
\begin{align}
\mathcal{P}:~~&\min_{\{\mathbf{r}_m\}_{m=1}^M, \mathbf{X}} ~ f\left(\{\mathbf{r}_m\}_{m=1}^M, \mathbf{X}\right)\\
&~~~\text { s.t. } 
 \mathbf{X} \in \mathcal{X},~\eqref{1}, \text{ and } \eqref{2} \nonumber,
\end{align}
where $f(\cdot)$ denotes a general utility function, $\mathbf{X}$ denotes the variables in the FPA systems, and $\mathcal{X}$ is the corresponding constraint set for $\mathbf{X}$. For example, in the MA-aided MIMO system of~\cite{ma_icc}, $f(\cdot)$ denotes the capacity, and $\mathbf{X}$ is the transmit covariance matrix. 

In general, if $f(\cdot)$ is a non-convex objective function or $\mathcal{X}$ is a non-convex set, the optimization for $\mathcal{P}$ is challenging even in the FPA systems. Furthermore, by introducing the MA's position variables $\{\mathbf{r}_m\}_{m=1}^M$, it is even more difficult to solve $\mathcal{P}$ because $\mathcal{X}$ and $\{\mathbf{r}_m\}_{m=1}^M$ are usually coupled in the objective function and constraints~\eqref{2} are non-convex. One existing approach regards $\mathcal{X}$ and $\{\mathbf{r}_m\}_{m=1}^M$ as two blocks and optimizes them through the alternating optimization~\cite{liyang1, liyang2}. However, when $f(\cdot)$ is non-convex with respect to $\{\mathbf{r}_m\}_{m=1}^M$, together with the non-convex constraints~\eqref{2}, it is challenging to tackle the subproblem with respect to $\{\mathbf{r}_m\}_{m=1}^M$ efficiently.

{\textcolor{black}{To separate the non-convex constraint~\eqref{2} from $f\left(\{\mathbf{r}_m\}_{m=1}^M, \mathbf{X}\right)$, we employ variable splitting by introducing $\{\mathbf{z}_m = \mathbf{r}_m\}_{m=1}^M$. Then, by replacing $\{\mathbf{r}_m\}_{m=1}^M$ in~\eqref{2} with  $\{\mathbf{z}_m\}_{m=1}^M$, and adding a penalty term $\rho\sum_{m=1}^M \left\|\mathbf{r}_m - \mathbf{z}_m\right\|_2^2$ to the objective function, $\mathcal{P}$ becomes:}
\begin{align}
&\mathcal{P}1:\min_{\{\mathbf{r}_m\}_{m=1}^M, \{\mathbf{z}_m\}_{m=1}^M, \mathbf{X}\in \mathcal{X}}~\nonumber\\
&\quad\quad\quad\quad\quad f\left(\{\mathbf{r}_m\}_{m=1}^M, \mathbf{X}\right)+\rho\sum_{m=1}^M \left\|\mathbf{r}_m - \mathbf{z}_m\right\|_2^2\\
&\quad\quad\quad\quad~~\text { s.t. } 
 \mathbf{r}_m \in \mathcal{C}, ~~\forall m =1,2, \ldots, M, \label{z1}\\
& \quad\quad\quad\quad\quad~~~~ \left\|\mathbf{z}_m-\mathbf{z}_l\right\|_2 \geq D, \nonumber \\
&\quad\quad\quad\quad\quad\quad~~~~\quad\quad\quad \forall m,l=1,2, \ldots, M, ~~ m \neq l, \label{z2}
\end{align}
where $\rho>0$ is the penalty factor\footnote{To choose a proper $\rho$, this paper initializes $\rho$ with a small value and then gradually increases $\rho$ to make the term $\sum_{m=1}^M \left\|\mathbf{r}_m - \mathbf{z}_m\right\|_2^2$ approach zero~\cite{qingfeng}.}. The problem $\mathcal{P}1$ can be solved by alternatively optimizing with respect to $\mathbf{X}$, $\{\mathbf{r}_m\}_{m=1}^M$ and $\{\mathbf{z}_m\}_{m=1}^M$, with details given in the following.
\subsubsection{Subproblem with respect to $\mathbf{X}$} Note that the penalty term $\rho\sum_{m=1}^M \left\|\mathbf{r}_m - \mathbf{z}_m\right\|_2^2$ does not depend on the value of $\mathbf{X}$. Thus, the subproblem with respect to $\mathbf{X}$ is given by
\begin{align}
\mathcal{P}\text{1-a}:~~&\min_{\mathbf{X}} ~~  f\left(\{\mathbf{r}_m\}_{m=1}^M, \mathbf{X}\right)\\
&\text { s.t. } 
\mathbf{X} \in \mathcal{X}.
\end{align}
\textcolor{black}{Note that the subproblem $\mathcal{P}\text{1-a}$ reduces to the general optimization form for conventional FPA systems, which can be solved by the existing approaches designed for the FPA systems.}
\subsubsection{Subproblem with respect to $\{\mathbf{r}_m\}_{m=1}^M$}
\begin{align} \label{p1_b}
\mathcal{P}\text{1-b}:~&\min_{\{\mathbf{r}_m \}_{m=1}^M}~  f\left(\{\mathbf{r}_m\}_{m=1}^M, \mathbf{X}\right)+\rho\sum_{m=1}^M \left\|\mathbf{r}_m - \mathbf{z}_m\right\|_2^2 \\
&~~\text { s.t. } 
  \eqref{z1}. \nonumber
\end{align}
Considering the existing works generally set $\mathcal{C}$ as a rectangular region~\cite{ma_icc,yang2024flexible}, the subproblem~$\mathcal{P}\text{1-b}$ can be efficiently solved to at least a stationary point by the projected gradient-based approach~\cite{nocedal2006numerical}.
\subsubsection{Subproblem with respect to $\{\mathbf{z}_m\}_{m=1}^M$} Since the variables $\{\mathbf{z}_m\}_{m=1}^M$ only appear in the penalty term, the resulting subproblem is given by
\begin{align} 
\mathcal{P}\text{1-c}:~~&\min_{\{\mathbf{z}_m\}_{m=1}^M} \sum_{m=1}^M \left\|\mathbf{z}_m - \mathbf{r}_m\right\|_2^2 \\
&~~\text { s.t. } 
  \eqref{z2}. \nonumber
\end{align}
Even though the problem $\mathcal{P}\text{1-c}$ is nonconvex due to the constraint $\left\|\mathbf{z}_m-\mathbf{z}_l\right\|_2 \geq D$, we present an efficient approach without any convex approximation. Specifically, we sequentially optimize~$\mathbf{z}_m$ with $\{\mathbf{z}_l\}_{l\neq m}$ being fixed and the $m$-th subproblem of $\mathcal{P}\text{1-c}$ is given by
\begin{align} \label{p1_c_m}
\mathcal{P}\text{1-c-m}:~~&\min_{{\mathbf{z}_m}}   \left\|\mathbf{z}_m - \mathbf{r}_m\right\|_2^2 \\
&\text { s.t. } 
\left\|\mathbf{z}_m-\mathbf{z}_l\right\|_2 \geq D, \nonumber \\
&\quad \quad \forall l=1,2, \ldots, M, \quad l \neq m. \label{z22}
\end{align}

\begin{figure*}
	\centering
 \subfigure[]{\label{111} 
		\includegraphics[width=1.5in]{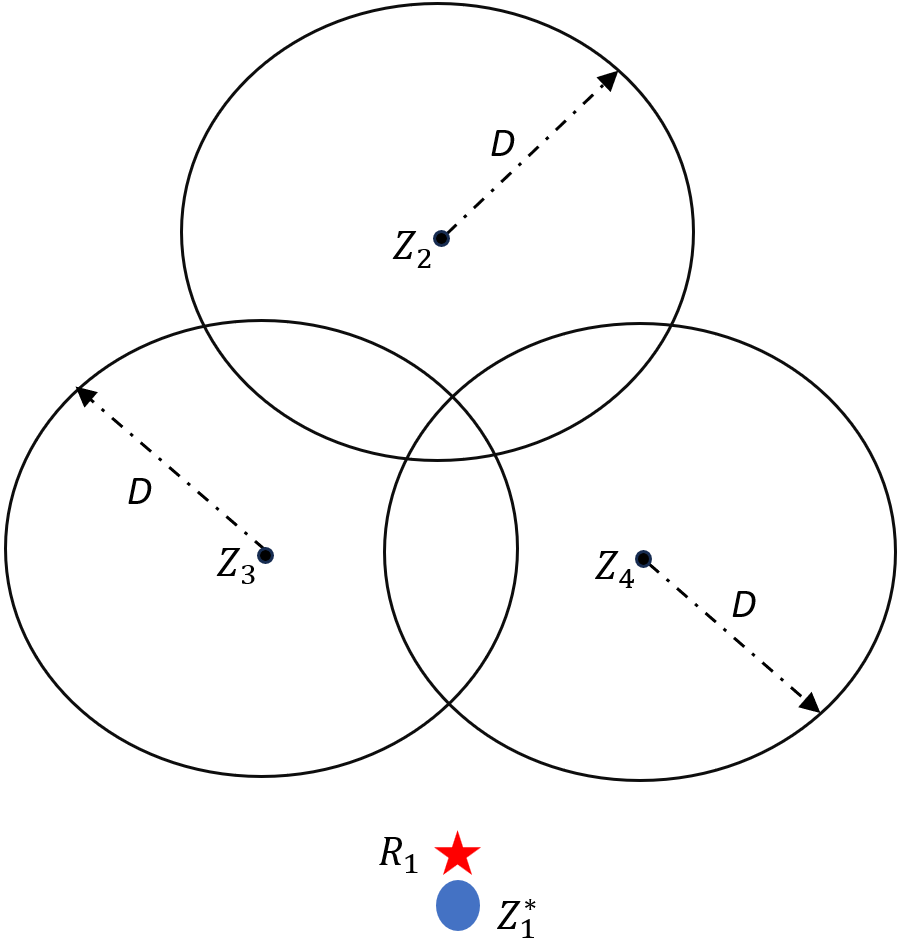}} 
   \subfigure[]{\label{222} 
		\includegraphics[width=1.5in]{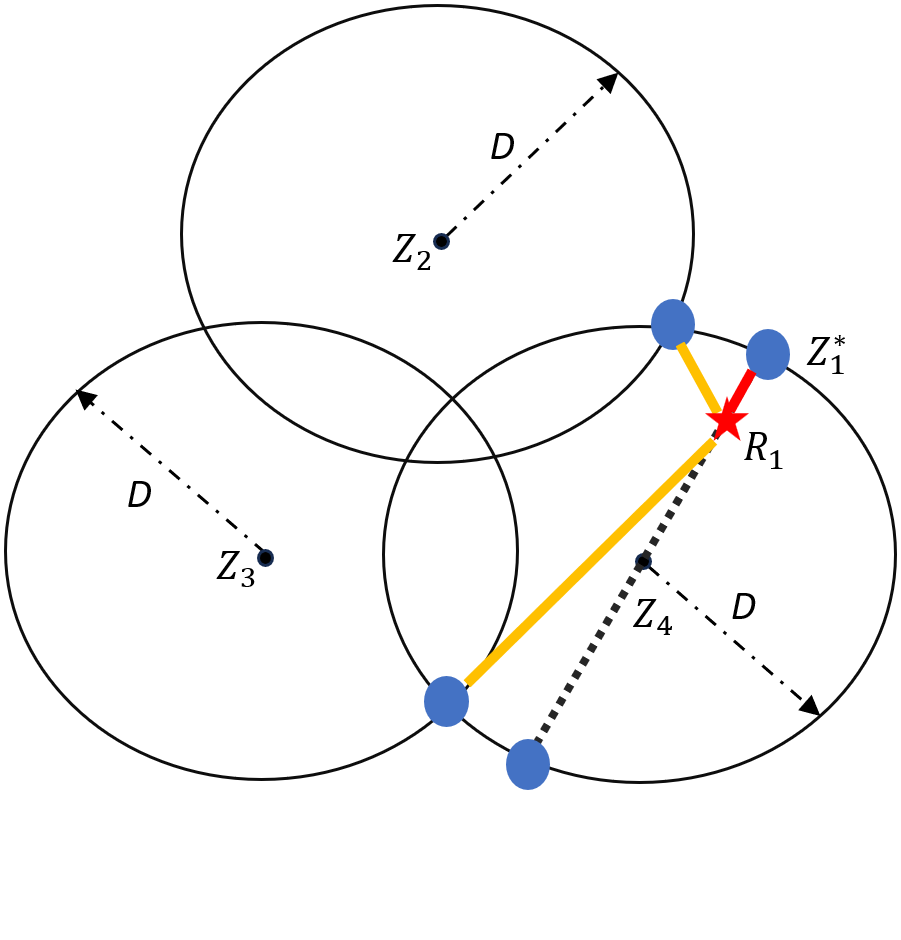}} 
	\subfigure[]{\label{333} 
		\includegraphics[width=1.5in]{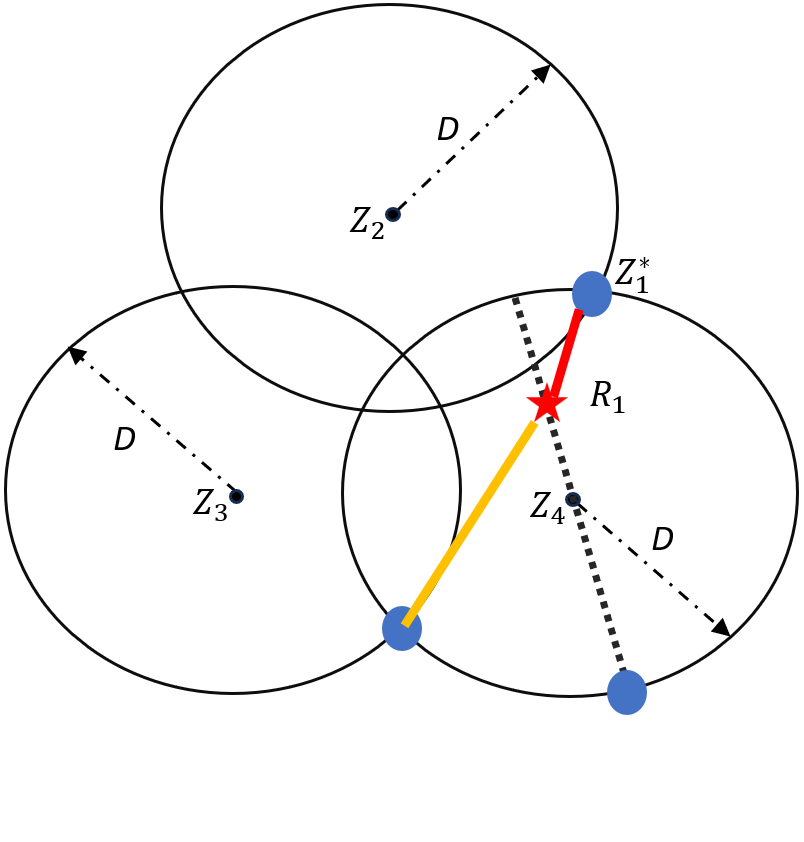}}  
	\subfigure[ ]{
		\label{666} 
		\includegraphics[width=1.5in]{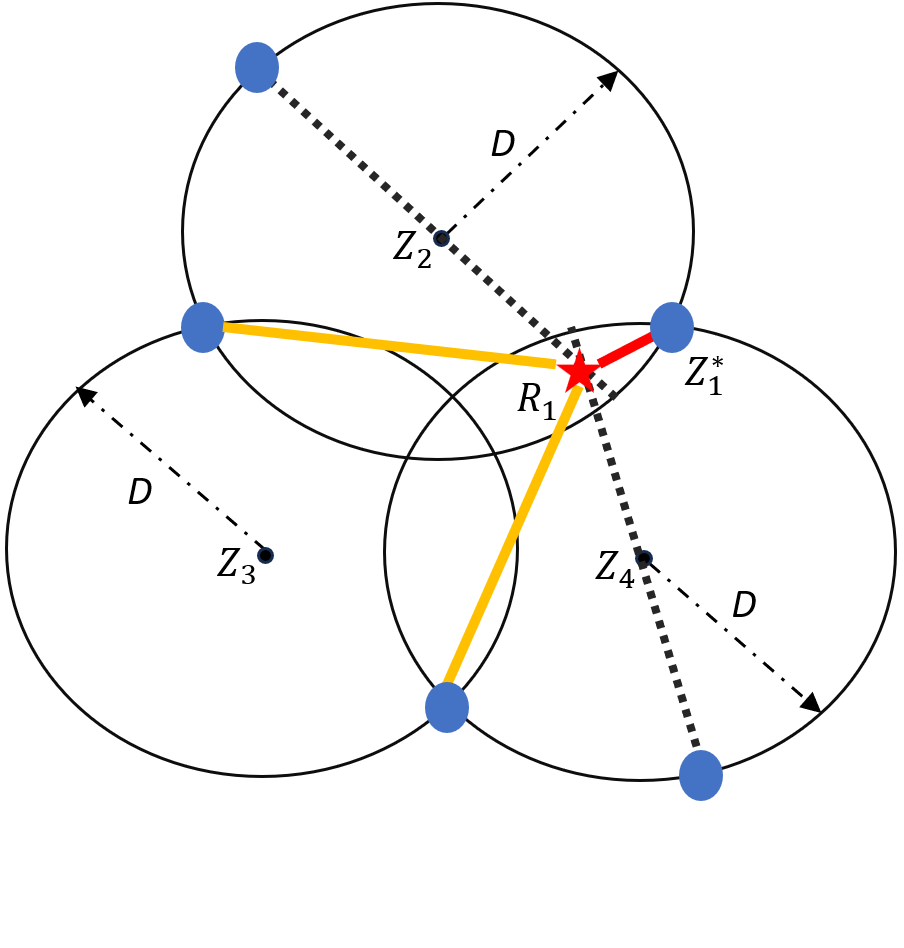}}
  	\caption{$m=1$ and $M=4$. The red lines denote the minimal distance between $\mathbf{r}_1$ and $\mathbf{z}_1$. (a) Illustration of case 1; (b) and (c) Illustration of case 2; (d) Illustration of case 3.} 
\end{figure*}

Before solving $\mathcal{P}\text{1-c-m}$, we define the following notations for the clear presentation. Let $\mathcal{L}$ denote the set of all $l$ that satisfy the constraint $\left\|\mathbf{r}_m-\mathbf{z}_l\right\|_2 < D$. The notation $\text{Circle}_{l}$ denotes the circle with the center $\mathbf{z}_l$ and radius~$D$. The notation $\mathcal{W}_l$ denotes the set of intersection points of $\text{Circle}_{l}$ with other circles (except for $\text{Circle}_{m}$), where these intersection points in the set need to satisfy the constraints~\eqref{z22}. The notation $\mathcal{U}_l$ denotes the set of intersection points of $\text{Circle}_{l}$ with the line passing through $\mathbf{z}_l$ and $\mathbf{r}_m$, where these points also need to satisfy the constraints~\eqref{z22}.

Then, we discuss the optimal solution of  $\mathbf{z}_m$ in $\mathcal{P}\text{1-c-m}$ into three cases.

\underline{(a)~$\mathcal{L}=\emptyset$:} The optimal $\mathbf{z}_m^{\star}$ is given by $\mathbf{z}_m^{\star} = \mathbf{r}_m^{\star}$ as shown in~Fig.~\ref{111}.

\underline{(b)~$|\mathcal{L}|=1$:} Under this case, we first need to find the only~$\mathbf{z}_l$, which satisfies $\left\|\mathbf{r}_m-\mathbf{z}_l\right\|_2 < D$. Then, the optimal $\mathbf{z}_m^{\star}$ belongs to the set $ \mathcal{W}_l \bigcup \mathcal{U}_l$,  which can be checked by contradiction. Therefore, as shown in Figs.~\ref{222} and~\ref{333}, the optimal $\mathbf{z}_m^{\star}$ is given by 
$\mathbf{z}_m^{\star} = \arg \min_{\mathbf{z}_m \in \mathcal{W}_l \bigcup \mathcal{U}_l} \left\|\mathbf{z}_m - \mathbf{r}_m\right\|_2^2$,
where the sets $\mathcal{W}_l$ and $\mathcal{U}_l$ can be obtained by using the geometry approaches.

\underline{(c)~$|\mathcal{L}|\geq 2$:} Under this case, as shown in Fig.~\ref{666}, the optimal $\mathbf{z}_m^{\star}$ is given by 
$\mathbf{z}_m^{\star} = \arg \min_{\mathbf{z}_m} \left\|\mathbf{z}_m - \mathbf{r}_m\right\|_2^2$ 
,where $\mathbf{z}_m \in \bigcup_{l=1}^{|\mathcal{L}|}\mathcal{V}_l $ and $\mathcal{V}_l \triangleq \mathcal{W}_l \bigcup \mathcal{U}_l$ . This can also be checked by contradiction.

The whole procedure for solving $\mathcal{P}1$ is summarized in Algorithm~\ref{a1}. \textcolor{black}{Due to space limitations, the complexity analysis of the Algorithm~\ref{a1} is left for future work}. Note that the stationary points are obtained for the subproblems with respect to $\{\mathbf{r}_m\}_{m=1}^M$ and $\{\mathbf{z}_m\}_{m=1}^M$. Once a stationary point can be obtained for the subproblem $\mathcal{P}\text{1-a}$, the convergence of Algorithm~\ref{a1} is guaranteed~\cite{razaviyayn2013unified}. The proposed optimization framework has two main advantages. On one hand, it can be observed that the non-convex antenna distance constraints have been transformed to the subproblem with respect to the auxiliary variables $\{\mathbf{z}_m\}_{m=1}^M$, which can be optimized without any approximation. On the other hand, the proposed optimization framework incorporates the conventional algorithms in FPA systems. This offers a streamlined procedure to solve the
MA based optimization problem. 

% \textcolor{blue}{The complexity of Algorithm~\ref{a1} for solving $\mathcal{P}\text{1-a}$ and $\mathcal{P}\text{1-b}$ is determined by the specific problem. Therefore, the discussion on the algorithm complexity only focuses on $\mathcal{P}\text{1-c}$. The worst-case complexity of this part is shown to be $\mathcal{O}(\gamma(M^2+N^2))$, where $\gamma$ denotes the maximum number of outer iterations of Algorithm~\ref{a1}.}

\begin{algorithm}[t] 
\caption{The Overall Algorithm for Handling $\mathcal{P}1$ } 
\begin{algorithmic}[1] \label{a1}
\STATE Initialize the optimization variables
\REPEAT
\STATE Update $\mathbf{X}$ based on $\mathcal{P}\text{1-a}$ using the approaches for the  FPA system.
\STATE Update $\{\mathbf{r}_m\}_{m=1}^M$ based on $\mathcal{P}\text{1-b}$ using the projected gradient-based approaches.
\REPEAT
\STATE \textbf{for} $m=1,\ldots,M$ \textbf{do}
\STATE ~~~~Update $\mathbf{z}_m$ by  solving problem $\mathcal{P}\text{1-c-m}$.
\STATE \textbf{end for}
\UNTIL Stopping criterion is satisfied.	
\UNTIL Stopping criterion is satisfied.	
\end{algorithmic}
\end{algorithm}
% \underline{Case (a) $\left\|\mathbf{r}_m-\mathbf{z}_l\right\|_2 \geq D, ~\forall l \neq m$:} The optimal solution $\mathbf{z}_m^{\star}$ is $\mathbf{r}_m$. 

% \underline{Case (b) $\left\|\mathbf{r}_m-\mathbf{z}_l\right\|_2 \geq D, ~\forall l \neq m$:}

\section{Case Studies} \label{case}
%In this section, we demonstrate the proposed optimization framework using two typical examples: capacity maximization and regularized zero-forcing precoding, respectively.

\subsection{Capacity Maximization for MA-Aided System~\cite{ma_icc}}

This case studies the channel capacity of a MIMO system with $M$ MAs at the BS and an $N$-antenna device. The received signal at the BS is given by
\begin{align}
\mathbf{y}\left(\{\mathbf{r}_m \triangleq [x_m,y_m]^T\}_{m=1}^M\right)=\mathbf{H}\left(\{\mathbf{r}_m\}_{m=1}^M\right) \mathbf{s}+\mathbf{z},
\end{align}
where $\mathbf{s} \in \mathbb{C}^{N}$ denotes the transmit signal and each elements of $\mathbf{z} \in \mathbb{C}^{M}$ are independent
and identically distributed (i.i.d.) Gaussian noise at the BS following $\mathcal{C}\mathcal{N}\left(0, \sigma_{\text{z}}^{2}\right)$ with $\sigma_{\text{z}}^{2}$ being the noise power. The notation $\mathbf{H}\left(\{\mathbf{r}_m\}_{m=1}^M\right)$ is the channel matrix from the device to BS, which is a function of $\{\mathbf{r}_m\}_{m=1}^M$. The field-response based channel model \textcolor{black}{provided in~\cite{MIMO_cap_cha_for_MA}} is
\begin{align}
\mathbf{H}\left(\{\mathbf{r}_m\}_{m=1}^M\right)\triangleq\left[\mathbf{b}\left(\mathbf{r}_1\right), \mathbf{b}\left(\mathbf{r}_2\right), \ldots, \mathbf{b}\left(\mathbf{r}_M\right)\right]^H \mathbf{\Sigma} \mathbf{G},
\end{align}
where $\mathbf{G} \in \mathbb{C}^{L_t \times N}$ is the field response matrix at the device side with $L_t$ being the number of transmitted paths, and the $(p, n)$-th element of $\mathbf{G}$ is given by $\exp\{j \pi \sin \theta_t^p \cos \phi_t^p(n-1)\}$ with $\theta_t^p$ and $\phi_t^p$ being the elevation and azimuth angles of
departure of the
$p$-th transmit path~$(p=1,2,\dots,L_t)$. The notation $\mathbf{b}(\mathbf{r}_m) \in \mathbb{C}^{L_r}$ is the field response vector of the $m$-th MA with $L_r$ being the number of received paths, and it is defined by
\begin{align}
\mathbf{b}(\mathbf{r}_m) \triangleq\left[\exp\left\{j \frac{2 \pi}{\lambda} \rho^1(\mathbf{r}_m)\right\}, \exp\left\{j \frac{2 \pi}{\lambda} \rho^2(\mathbf{r}_m)\right\}, \right.\nonumber \\
\left.\ldots, \exp\left\{j \frac{2 \pi}{\lambda} \rho^{L_r}(\mathbf{r}_m)\right\}\right]^T,
\end{align}
where $\lambda$ is the wavelength and $\rho^q(\mathbf{r}_m)\triangleq x_m \sin \theta_r^q \cos \phi_r^q+$ $y_m \cos \theta_r^q$ with $\theta_r^q$ and $\phi_r^q$ being the elevation and azimuth angles of
arrival of the
$q$-th receive path~$(q=1,2,\dots,L_r)$. The notation~$\mathbf{\Sigma} \in \mathbb{C}^{L_r \times L_t}$ denotes the response between the transmit paths and receive paths.

Assume that perfect channel state information is available at both the device and BS, the problem of capacity maximization is formulated as
\begin{align}
&\mathcal{P}\text {(A1):}~~\min_{\{\mathbf{r}_m\}_{m=1}^M, \mathbf{Q}} \quad \nonumber \\
& -\log _2 \operatorname{det}\left(\mathbf{I}_M+\frac{1}{\sigma^2} \mathbf{H}\left(\{\mathbf{r}_m\}_{m=1}^M\right) \mathbf{Q H}\left(\{\mathbf{r}_m\}_{m=1}^M\right)^H\right)  \\
& \quad\quad\quad\quad\text { s.t. } 
\operatorname{Tr}(\mathbf{Q}) \leq P_{\text{max}},~\mathbf{Q} \succeq \mathbf{0},  ~\eqref{1}, \text{ and   } \eqref{2}, \nonumber
\end{align}
where $\mathbf{Q} \triangleq \mathbb{E}\left\{\mathbf{s} \mathbf{s}^H\right\} $ denotes the transmit covariance matrix and $P_{\text{max}}$ is the maximum transmit power for the device. One existing approach~\cite{ma_icc} regards $\mathbf{Q}$ and $\{\mathbf{r}_m\}_{m=1}^M$ as two blocks and optimizes them  through the alternating optimization. However, given fixed $\mathbf{Q}$, the objective function and the constraint~\eqref{2} with respect to $\{\mathbf{r}_m\}_{m=1}^M$ are both nonconvex. To this end, [1] leverages SCA to both objective function and constraints. In the following section, we demonstrate that the proposed general optimization framework can be applied to effectively solve problem~$\mathcal{P}\text {(A1)}$ without any approximation.

By applying the proposed framework, problem $\mathcal{P}\text {(A1)}$ becomes
\begin{align}
&\mathcal{P}\text {(A2):}~~\min_{\{\mathbf{r}_m\}_{m=1}^M, \{\mathbf{z}_m\}_{m=1}^M, \mathbf{Q}} \quad & \nonumber \\
&-\log _2 \operatorname{det}\left(\mathbf{I}_M+\frac{1}{\sigma^2} \mathbf{H} \mathbf{Q H}^H\right) +\rho\sum_{m=1}^M \left\|\mathbf{r}_m - \mathbf{z}_m\right\|_2^2 \label{o}\\ 
&~~~~ \text { s.t. } 
\operatorname{Tr}(\mathbf{Q}) \leq P_{\text{max}},~\mathbf{Q} \succeq \mathbf{0},~\eqref{z1},\text{ and}~ \eqref{z2}. \nonumber
\end{align}
Under the proposed framework, the problem $\mathcal{P}\text {(A2)}$ is solved by alternatively
 optimizing the blocks $\mathbf{Q}$, $\{\mathbf{r}_m\}_{m=1}^M$, and $\{\mathbf{z}_m\}_{m=1}^M$. Since the optimization of $\{\mathbf{z}_m\}_{m=1}^M$ does not rely on the specific problem and has been solved in Section~\ref{framework}, we only present the details about the optimization of $\mathbf{Q}$ and $\{\mathbf{r}_m\}_{m=1}^M$.
 
\underline{(a)~Subproblem with respect to $\mathbf{Q}$:} \\
By denoting $\mathbf{H}\left(\{\mathbf{r}_m\}_{m=1}^M\right)=\tilde{\mathbf{U}} \tilde{\boldsymbol{\Lambda}} \tilde{\mathbf{V}}^H$ as the truncated singular
value decomposition of $\mathbf{H}\left(\{\mathbf{r}_m\}_{m=1}^M\right)$, where $\tilde{\mathbf{U}} \in \mathbb{C}^{M \times S}$, $\tilde{\boldsymbol{\Lambda}} \in \mathbb{C}^{S \times S}$, $\tilde{\mathbf{V}} \in \mathbb{C}^{N \times S}$, and $S = \text{rank}\left(\mathbf{H}\left(\{\mathbf{r}_m\}_{m=1}^M\right)\right)$, the optimal $\mathbf{Q}^{\star}$ is given by
\begin{align} \label{28}
\mathbf{Q}^{\star}=\tilde{\mathbf{V}} \operatorname{diag}\left(\left[p_1^{\star}, p_2^{\star}, \ldots, p_S^{\star}\right]\right) \tilde{\mathbf{V}}^H,
\end{align}
where $p_s^{\star}=\max \left(1 / p_0-\sigma^2 / \tilde{\boldsymbol{\Lambda}}[s, s]^2, 0\right)$ with $p_0$ satisfying $\sum_{s=1}^S p_s^{\star}=P_{\text{max}}$. The details of deriving~\eqref{28} can be checked in~\cite{ma_icc}.

\underline{(b)~Subproblem with respect to $\{\mathbf{r}_m\}_{m=1}^M$:} \\
Consider the objective function~\eqref{o} is differentiable with respect to $\{\mathbf{r}_m\}_{m=1}^M$, we applied the projected gradient approach to optimize $\{\mathbf{r}_m\}_{m=1}^M$. For each iteration $i$,  $\{\mathbf{r}_m^{(i)}\}_{m=1}^M$ is updated by
\begin{align}
\{\mathbf{r}_m^{(i)}\}_{m=1}^M = \mathcal{P}_{\mathcal{C}}\left\{\{\mathbf{r}_m^{(i-1)}\}_{m=1}^M - \eta^{(i-1)} \nabla G\left(\{\mathbf{r}_m^{(i-1)}\}_{m=1}^M\right)\right\}, \nonumber
\end{align}
where $G\left(\{\mathbf{r}_m^{(i-1)}\}_{m=1}^M\right) \triangleq -\log _2 \operatorname{det}\left(\mathbf{I}_M+\frac{1}{\sigma^2} \mathbf{H} \mathbf{Q H}^H\right) +\rho\sum_{m=1}^M \left\|\mathbf{r}_m - \mathbf{z}_m\right\|_2^2$, and its gradient $\nabla G\left(\{\mathbf{r}_m^{(i-1)}\}_{m=1}^M\right)$ can be obtained by employing the auto-differentiation mechanism of PyTorch. The notation $\mathcal{P}_{\mathcal{C}}\left\{\cdot\right\}$ denotes the projection on the feasible set $\mathcal{C}$.

\subsection{Regularized Zero-forcing Precoding for Multi-user MA-Aided System~\cite{yang2024flexible}}

This case studies the regularized zero-forcing (RZF) precoding scheme of the multi-user MISO system with $M$ MAs at the BS and $K$ single-antenna devices. The received signal at the $k$-th device is given by 
\begin{align}
\ y_k\left(\{\mathbf{r}_m\}_{m=1}^M\right)=\mathbf{h}_k\left(\{\mathbf{r}_m\}_{m=1}^M\right)^H \mathbf{F} \mathbf{s}+ z_k,
\end{align}
where $\mathbf{s} \in \mathbb{C}^{K \times 1}$ represents the data streams for all $K$ devices with $ \mathbb{E}\left\{\mathbf{s} \mathbf{s}^H\right\}=\mathbf{I}_K$, $\mathbf{F} \triangleq [\mathbf{f}_1,\mathbf{f}_2,...,\mathbf{f}_K] \in \mathbb{C}^{M \times K}$ is the precoding matrix, $z_k \sim \mathcal{C}\mathcal{N}\left(0, \sigma^{2}_k\right)$ denotes the Gaussian noise, and $\mathbf{h}_k\left(\{\mathbf{r}_m\}_{m=1}^M\right)^H \in \mathbb{C}^{M}$ is the channel vector of the $k$-th device which is modelled following~\cite[eq.~(3)]{yang2024flexible}. Then, the RZF problem is formulated as
\begin{align}
\mathcal{P}\text {(B1):}~&\min_{\{\mathbf{r}_m\}_{m=1}^M, \mathbf{F}} ~ \left\|\mathbf{I}_K-\mathbf{H} \left(\{\mathbf{r}_m\}_{m=1}^M\right)\mathbf{F}\right\|_F^2+\alpha\|\mathbf{F}\|_F^2 \nonumber\\
&~~~~ \text { s.t. } 
\eqref{1} \text{   and   } \eqref{2} \nonumber,
\end{align}
where $\alpha$ is a hyperparameter controlling the flexibility of precoding. One existing approach~\cite{yang2024flexible} equivalently transforms the original problem $\mathcal{P}\text {(B1)}$ to the sparse optimization and leverages the CS based approach to tackle the transformed problem. However, this approach introduces the additional zero-norm constraint, which is challenging to tackle and leads to the performance loss. 

Next, we demonstrate the use of proposed general optimization framework to effectively solve the problem $\mathcal{P}\text {(B1)}$.
By applying the proposed framework, problem $\mathcal{P}\text {(B1)}$ becomes
\begin{align}
&\mathcal{P}\text {(B2):}~~\min_{\{\mathbf{r}_m\}_{m=1}^M, \{\mathbf{z}_m\}_{m=1}^M, \mathbf{F}} \quad & \nonumber \\
&\left\|\mathbf{I}_K-\mathbf{H} \left(\{\mathbf{r}_m\}_{m=1}^M\right)\mathbf{F}\right\|_F^2+\alpha\|\mathbf{F}\|_F^2 +\rho\sum_{m=1}^M \left\|\mathbf{r}_m - \mathbf{z}_m\right\|_2^2 \label{o1}\\ 
&~~~~~~~~~~~~~ \text { s.t. } 
\eqref{z1} ~\text{and}~ \eqref{z2}.\nonumber
\end{align}
Following the similar logic as in the case of capacity maximization, when fixing $\{\mathbf{r}_m\}_{m=1}^M$ and $\{\mathbf{z}_m\}_{m=1}^M$, the optimal $\mathbf{F}^{\star}$ is given by $\mathbf{F}^{\star} = \left(\mathbf{H}^H \mathbf{H}+\alpha \mathbf{I}\right)^{-1} \mathbf{H}^H$. Together with the fact that the objective function~\eqref{o1} is differentiable with respect to $\{\mathbf{r}_m\}_{m=1}^M$, the problem $\mathcal{P}\text {(B2)}$ can be solved by alternatively
optimizing the blocks $\mathbf{F}$, $\{\mathbf{r}_m\}_{m=1}^M$, and $\{\mathbf{z}_m\}_{m=1}^M$.

\section{Numerical Results}
In this section, we demonstrate the superiority of the proposed optimization framework on the two examples introduced in Section~\ref{case} via simulations, where the stopping criterion of Algorithm 1 is the relative variation of the objective value being no more than $10^{-3}$. In both cases, we consider the BS is equipped with $M=4$ MAs and the MA region $\mathcal{C}$ is set as an $A \times A$ square area. The minimum distance between MAs is set as $D=\lambda/2$. \textcolor{black}{The initial value of $\rho$ is 5 and it increases by 1.2 times in each iteration.}

\subsection{Capacity Maximization for MA-Aided System}

\begin{figure}[t!]
\begin{center}
  \includegraphics[width=0.33\textwidth]{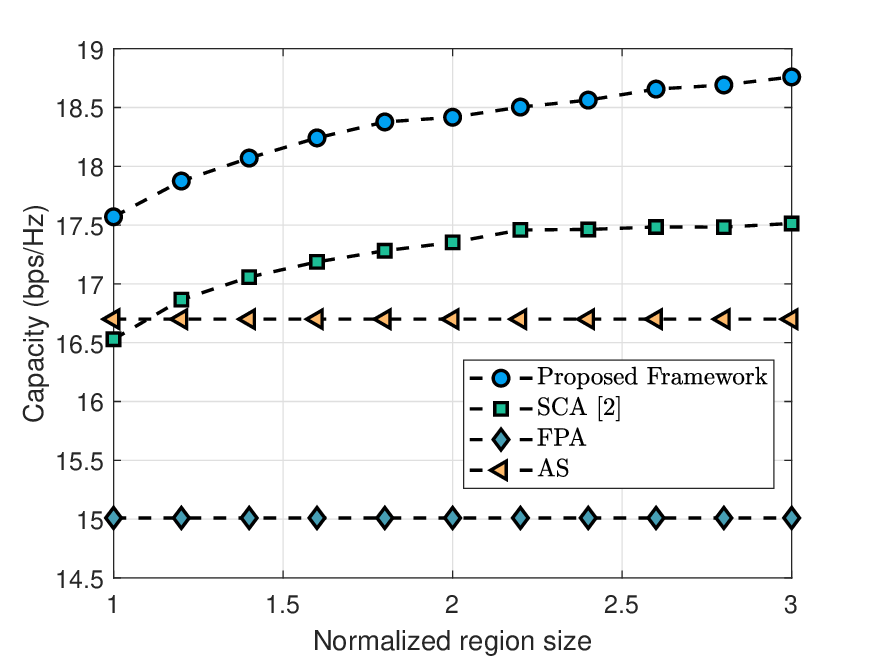}
  \caption{Capacity versus normalized receive region size. The parameters of the channel model are the same as those in~\cite{ma_icc} with $L_t=L_r=10$.}\label{f1}
\end{center}
\end{figure}

In this case, we consider a MIMO system with a $4$-antenna uniform linear array device. For comparison, we provide the following baselines:
\begin{itemize}
    \item FPA: Both the BS and device have $4$ fixed antennas that are spaced by~$\lambda/2$ \textcolor{black}{in the square region.}
    \item Antenna Selection (AS): Both sides are equipped with $8$ fixed antennas, \textcolor{black}{which are distributed in the square region} and spaced by $\lambda/2$. The algorithm selects half of these antennas to work via an exhaustive search.
    \item SCA~\cite{ma_icc}: The BS is equipped with 4 MAs and the device is equipped with 4 fixed antennas. By leveraging SCA, It iteratively optimizes
    the transmit covariance matrix and the position of each MA with the other variables being fixed.
\end{itemize}

In Fig.~\ref{f1}, we demonstrate the channel capacity versus the normalized region size $A/\lambda$ for the proposed optimization framework and the baselines. It is observed that the proposed optimization framework outperforms all baselines. Specifically, compared with FPA and AS, the proposed optimization framework flexibly optimizes the positions of the MAs in a continuous area, thus enabling to exploit more spatial degrees of
freedom for enhancing the channel capacity. Compared with SCA in~\cite{ma_icc}, the proposed framework does not introduce any approximation and hence achieves better performance.

\subsection{Regularized Zero-forcing Precoding for Multi-user MA-Aided System}

\begin{figure}[t!]
\begin{center}
  \includegraphics[width=0.33\textwidth]{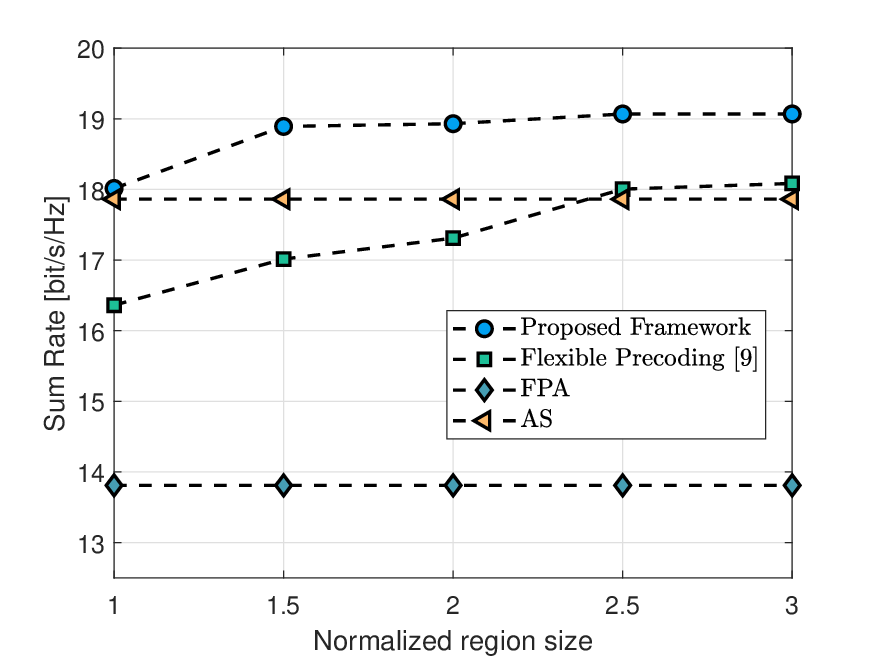}
  \caption{The sum rate versus normalized receive region size with $\alpha = 6$. The parameters of the channel model are the same as those in~\cite{yang2024flexible}. }\label{f2}
\end{center}
\end{figure}

In this case, we consider the multi-user MISO system with $K=4$ single-antenna devices. For comparison, except for the baselines FPS and AS, we additional provide the following baseline:
\begin{itemize}
    \item Flexible Precoding~\cite{yang2024flexible}: It introduces the zero-norm constraint and makes use of the CS based approach to design the precoder.
\end{itemize}

In Fig.~\ref{f2}, we illustrate the sum rate versus the normalized region size $A/\lambda$ for the proposed optimization framework and the baselines. It is also seen that the proposed optimization framework outperforms all baselines. It is because  the proposed optimization framework does not introduce any non-convex or non-smooth constraints (i.e., zero-norm constraint) and hence it achieves satisfactory performance.

\section{Conclusion}

In this paper, we proposed a general optimization framework for the MA-aided system under the non-convex antenna distance constraints. In particular, we introduced auxiliary variables to separate the non-convex antenna distance constraints from the objective function. Then, the resulting problem was solved under the alternating optimization framework. We applied the proposed framework to two examples: capacity maximization and regularized zero-forcing precoding. Numerical results confirmed the effectiveness of the proposed optimization framework over other state-of-the-art methods. Moreover, the proposed framework can be extended to various communication scenarios in the MA-aided system.

\bibliographystyle{IEEEtran}
\bibliography{ref}

\end{document}